\documentclass[aps,pra,showpacs,twocolumn]{revtex4-1}
\usepackage{graphicx,psfrag,amsmath,amssymb,amsfonts,latexsym,color,dcolumn}
\begin{document}
\date{}
\title{Radiation from a moving planar dipole layer: patch
potentials vs dynamical Casimir effect}
\author{C\'esar D. Fosco}
\author{Francisco D. Mazzitelli}
\affiliation{ Centro At\'omico Bariloche,
Comisi\'on Nacional de Energ\'\i a At\'omica,
R8402AGP Bariloche, Argentina}
\affiliation{ Instituto Balseiro,
Universidad Nacional de Cuyo,
R8402AGP Bariloche, Argentina}
\begin{abstract} 
We study the classical electromagnetic radiation due to the presence of a
dipole layer on a plane that performs a bounded motion along its normal
direction, to the first non-trivial order in the amplitude of that motion.
We show that the total emitted power may be written in terms of the dipole
layer autocorrelation function.  We then apply the general expression for
the emitted power to cases where the dipole layer models the presence of
patch potentials, comparing the magnitude of the emitted radiation with
that coming from the quantum vacuum in the presence of a moving perfect
conductor (dynamical Casimir effect). 
\end{abstract}
\maketitle
\section{Introduction}\label{sec:intro}

Due to the unavoidable presence of imperfections, impurities, and even
spatial variations in its chemical composition, the surface of a real metal
is not a perfect equipotential.  In other words, those effects manifest
themselves in the existence of an electrostatic potential, with a non
trivial space dependence, on the metallic surface \cite{Lang}. These
residual ``patch potentials" may even produce forces between different metallic
surfaces,  that become relevant for high sensitivity experiments, in
many areas of Physics \cite{experiments}.  In particular, they can be
crucial when determining the vacuum forces of quantum electromagnetic (EM)
origin between neutral objects (static Casimir effect) and in experiments
looking for modifications to the gravitational inverse-square law
\cite{experiments, Speake,FLM2013,DD2013}.  In those experimental setups, the
existence of patch potentials results in the presence of a force,
electrostatic in origin, and the theoretical goal is then to have a proper,
and hopefully simple, way to quantify it. Ideally, one should be able to
disentangle its contribution in any experimental attempt to determine
forces which are of a different nature. For static situations, one is
usually able to derive general expressions of the effects due to the patch
potentials in terms of just their autocorrelation functions. 

In this paper, we quantify an effect due to the presence of patch potentials in a
rather different, complementary situation: namely, we consider a metallic
object which undergoes accelerated motion, and study the resulting EM
radiation.
The physical reason to expect such radiation becomes clear when one recalls
that patch potentials can be thought of as due to the existence of a
(space-dependent) dipole layer on the surface of an otherwise neutral body
\cite{Speake}. Therefore, when accelerated, the moving dipole layer shall emit
radiation.  Our main goal here is to evaluate the power emitted by a flat
surface containing patch potentials, in terms of the acceleration of the
plane and the characteristics of the patch potentials.

One of the motivations that lead us to this calculation is to make a
quantitative comparison between the magnitudes of this classical effect with the power
emitted by an accelerated ideal, perfectly conducting mirror. This ``motion induced
radiation" or ``dynamical Casimir effect" (DCE) \cite{reviews dce} is a purely quantum
effect, that, as we will see, could also be interpreted as coming from a
moving dipole layer with an {\em ad hoc\/} autocorrelation function.  

The radiation field of a single time-dependent dipole at rest is a classical
problem, described in almost all texts on classical electrodynamics.  
The radiation field of a moving dipole, however, is not so widely known,
although it has already been investigated in the sixties \cite{movingdip}.  
To our knowledge, the spectrum of classical electromagnetic (EM)
radiation due to the presence of a {\em dipole layer\/} on a plane that moves
rigidly has not been computed before. Therefore, since it is a crucial
ingredient for our study, we present this calculation in
Section~\ref{sec:radiation}. 
As we shall show, there is a rather simple formula for the spectral density
associated to the radiated energy, in terms of a two point function that
describes the correlation of the dipoles at different points of the
surface. Using the model of Ref. \cite{Speake}, this leads immediately to an
expression for the emitted power by moving patch potentials. In this section, 
we also comment on the case of time dependent dipole layers.
In Section~\ref{sec:examples}, we compute the spectral density for the
particular autocorrelations
previously used in the literature to describe patch potentials, and make a 
comparison between  the classical emitted power and the DCE. 
Section~\ref{sec:conc} contains the conclusions of our work.  

\section{Radiation}\label{sec:radiation}

In this section we
evaluate the classical EM radiation due to the presence of a
dipole layer on a plane that moves rigidly along the direction defined by
its normal (we use CGS-Gaussian units throughout). 
The instantaneous position of the plane may  be defined in
terms of a single function $q(t)$ such that $x_3 = q(t)$. Here, $x_3$ is
one of the three Cartesian coordinates $(x_1,x_2,x_3)$, for which we shall
also use the notation ${\mathbf x_\parallel} \equiv (x_1,x_2)$.
The dipole layer density $D$, shall then be a function of the two
coordinates parallel to the plane, namely, with the notation just
introduced, $D = D({\mathbf x}_\parallel)$. 

To proceed to the calculation of the emitted radiation, we need the charge
and current densities $\rho$ and ${\mathbf j}$, which for the system we are
considering are given by:
\begin{eqnarray}\label{eq:sources}
\rho({\mathbf x},t) &=& - D({\mathbf x}_\parallel) \, \delta'(x_3-q(t)) \nonumber\\
{\mathbf j}({\mathbf x},t) &=& - D({\mathbf x}_\parallel)
\, \delta'(x_3-q(t)) \, {\dot q}(t)\,  \mathbf{\hat e_3} \;,
\end{eqnarray}
where $\mathbf{\hat e_3}$ is the unit vector along the direction of motion.

Finally, we shall assume that the motion is bounded, namely, that there is
a length $l$ such that $|q(t)| \leq l$, $\forall t$.

To determine the radiated power, we introduce the retarded potentials, and
use the Lorentz gauge fixing condition, obtaining for the potentials the
inhomogeneous wave equations: 
\begin{eqnarray}
\phi({\mathbf x},t) &=& \int d^3x'dt' \, G({\mathbf x},t;{\mathbf x'}, t') 
\, \rho({\mathbf x'},t') \nonumber\\
{\mathbf A}({\mathbf x},t) &=& \frac{1}{c} \, \int d^3x'dt' 
G({\mathbf x},t;{\mathbf x'}, t')
{\mathbf j}({\mathbf x'},t') \;,
\end{eqnarray}
where $G$ denotes the retarded Green's function for the wave equation, 
which satisfies:
\begin{equation}
(c^{-2} \partial_t^2 - \nabla_{\mathbf x}^2) G({\mathbf x},t;{\mathbf
x'}, t') \;=\; 4\pi \, \delta({\mathbf x}-{\mathbf x'}) \, \delta(t-t') \;.
\end{equation}
A more explicit expression may be obtained by introducing the Fourier
transformation: 
\begin{eqnarray}
G({\mathbf x},t;{\mathbf x'}, t') &=& \int \frac{d\omega}{2\pi}
\frac{d^3{\mathbf k}}{(2\pi)^3} \; e^{ -i \omega (t -t') + i {\mathbf k}
\cdot ({\mathbf x}-{\mathbf x'})} \nonumber\\
&\times & {\widetilde G}({\mathbf k}_\parallel,k_3,\omega)
\;,
\end{eqnarray}
with
\begin{equation}
{\widetilde G}({\mathbf k}_\parallel,k_3,\omega)=\frac{4\pi}{{\mathbf
k}_\parallel^2+k_3^2-(\frac{\omega}{c}+i\eta)^2} \;.
\end{equation}

The next step in the derivation of the emitted power is the introduction of
the  Poynting vector, ${\mathbf S}=\frac{c}{4\pi}{\mathbf E}\times {\mathbf
B}$ where:
\begin{eqnarray}
{\mathbf E}&=& -\nabla\phi-\frac{1}{c}\frac{\partial}{\partial t}{\mathbf A}\nonumber\\
{\mathbf B}&=&\nabla\times{\mathbf A} \;.
\end{eqnarray}
Of course, for an arbitrary dipole layer, the radiation flux may have a
rather cumbersome spatial dependence on the details of the layer. 
On the other hand, one is presumably more interested in the global effect,
namely, the average flux of energy, since the spatial dependence of that
flux is hardly detectable. 
The geometry of the system suggests to
evaluate the total flux of radiated energy due to the moving plane. It is
then convenient to evaluate the third component of ${\mathbf S}$, on a
constant-$x_3$ plane,  far from the region where the plane moves ($|x_3| > l$).

We see that the third component of ${\mathbf S}$ may be written as follows:
\begin{equation}\label{eq:s3}
S_3 \;=\; \frac{c}{4\pi} \, \epsilon_{ij} \, E_i B_j \;,
\end{equation}
where the indices $i$, $j$ shall be assumed, from now on, to run from $1$
to $2$. 
Since ${\mathbf A}$ points in the $ \mathbf{\hat e_3}$ direction, we see
that the components of the electric and magnetic field relevant to
the calculation of (\ref{eq:s3}) are given by:
\begin{equation}
E_i \,=\, - \partial_i \phi \;,\;\; B_i \,=\, \epsilon_{ij} \partial_j A_3
\;.
\end{equation}
Thus $S_3$ becomes,
\begin{equation}
S_3 \,=\,\frac{c}{4\pi} \, \partial_j \phi \, \partial_j A_3 \;. 
\end{equation}

The total flux of energy through one such plane shall be, in general, a
divergent quantity, something that may be dealt with by dividing it by the
total area. Besides, it is also convenient to calculate the total radiated
energy since, when written in terms of the Fourier transforms of the
time-dependent functions, it will allow us to extract the spectral density
of radiation. 

Thus, the (average) radiated energy per unit area through a constant-$x_3$
plane becomes:
\begin{equation}
U_{rad}(x_3)\;=\; \frac{1}{L^2}\int dt\int d^2{\mathbf x}_{\parallel} \,
S_3({\mathbf x}_\parallel,x_3,t)
\end{equation}
where $L^2$ is the area of the plane, assumed temporarily large and finite,
but an $L \to \infty$ limit at the end is assumed. 
We expect it to be independent of $x_3$ far from the planes.

After some algebra, we see that, to second order in $q(t)$, $U_{rad}(x_3)$ may
be written as follows:
\begin{eqnarray}
U_{rad}(x_3) &=& -4\pi \,\int
\frac{d^2k_\parallel}{4\pi^2}\frac{d\omega}{2\pi}\frac{dk_3}{2\pi}\frac{dp_3}{2\pi}
\nonumber\\
&\times& \Big\{ \frac{|{\mathbf k_\parallel}|^2 k_3^2 p_3 \omega}{\big[ {\mathbf
k}_\parallel^2 + k_3^2 -(\frac{\omega}{c} + i \eta)^2 \big]
\big[ {\mathbf k}_\parallel^2 + p_3^2 -(\frac{\omega}{c} - i \eta)^2 \big]}
\nonumber\\
&& \widetilde{\Omega}({\mathbf k}_\parallel) \; |\tilde{q}(\omega)|^2
e^{ix_3(k_3+p_3)} \Big\}\;,
\end{eqnarray}
where we have introduced the Fourier transform of the dipole layer autocorrelation function 
\begin{equation}
\Omega({\mathbf x}_\parallel)\,=\, \frac{1}{L^2} \, 
\int d^2 y_\parallel  D({\mathbf y}_\parallel) D({\mathbf
x}_\parallel+{\mathbf y}_\parallel)\;.
\end{equation}
It is worth noting that, in natural ($\hbar =1$ and $c=1$) units,
${\widetilde \Omega}$ is a dimensionless quantity.
We mention at this point that a similar expression to the one above could
have been obtained if one had a random patch potential distribution, with a
translation invariant stochastic correlation. Namely, even without evaluating
the average over a constant-$x_3$ plane, the translation invariance of the
system does produce an entirely analogous expression to the one above, now
interpreting $\Omega$ as the result of an average with a statistical
weight.

We then evaluate the integrals over $k_3$ and $p_3$, which can be
performed, for example, by using Cauchy's theorem in a rather
straightforward way, obtaining a result that, contains both convection and
radiation terms. The latter are, for $x_3> 0$, independent of $x_3$. On the
other hand, an evaluation of $U_{rad}(-x_3)$, the average energy flux
through a plane symmetrically located with respect to $x_3=0$, yields the
same result as $U_{rad}(x_3)$.  

Introducing $U_{rad}$, the total radiated energy per unit area:
\begin{equation}
U_{rad} \,=\, U_{rad}(x_3) \,+\,  U_{rad}(-x_3) \,=\, 2  U_{rad}(x_3) \;,
\end{equation}
which, moreover, may be conveniently written as follows:
\begin{equation}\label{eq:desp0}
U_{rad} \,=\, \int_0^\infty \frac{d\omega}{2\pi} \, {\mathcal P}({\omega}) \;,
\end{equation}
where the spectral density ${\mathcal P}({\omega})$ is:
\begin{equation}\label{eq:desp}
{\mathcal P}(\omega)\, = \, \vert\omega\vert\vert\tilde
q(\omega)\vert^2\int_0^{\omega/c}dk_\parallel\,
k_\parallel^3 \, \widetilde{\Omega}(k_\parallel) \,
\sqrt{(\frac{\omega}{c})^2-k_\parallel^2}
\end{equation}
where we have assumed the autocorrelation function to be isotropic.

Equations (\ref{eq:desp0}), (\ref{eq:desp}) constitute the main result of
this section, namely, a rather general and compact expression for the spectral
density of emitted energy in terms of the two main ingredients that
characterize the system: its motion $\tilde{q}(\omega)$ and the
autocorrelation function of the dipole layer density.

Assuming that the correlation length is much
smaller than $c/\omega$,  we can approximate 
$\widetilde\Omega(k_\parallel)\simeq \widetilde\Omega(0)$. With this approximation
\begin{equation}
U_{rad}\simeq\frac{2}{15}\widetilde\Omega(0)\int dt\, \dot a^2
\end{equation}
where $a$ is the acceleration.  From this equation, one can show that
the radiation reaction force per unit area on the dipole layer is
\begin{equation}
f_{rad}\simeq\frac{2}{15}\widetilde\Omega(0)\dddot a
\end{equation}
Notably, in this approximation the results coincide 
with those  of a
single moving dipole $d$ with $d^2\equiv\widetilde\Omega(0) c^5$
\cite{movingdip,movingdip2}. 

Up to here we considered a dipole layer. In order to describe an imperfect
conductor with patch potentials, following Ref.\cite{Speake} we can
consider a dipole layer close to a grounded perfect conductor. For this
configuration, and for non relativistic motion of the mirror, the radiated
power can be computed using the method of images: in the presence of the
grounded conductor, the dipole density is increased by a factor 2, and
therefore the spectral density by a factor 4. 

Finally, we consider  the case of time dependent dipole layers, namely,
$D = D({\mathbf x_\parallel}, t)$. This time dependence may be produced by an external agent,
or may describe intrinsic fluctuations of the material. The generalization of (\ref{eq:sources})
to this situation is:
\begin{eqnarray}\label{eq:sourcest}
\rho({\mathbf x},t) &=& - D({\mathbf x}_\parallel,t) \, \delta'(x_3-q(t)) \nonumber\\
{\mathbf j}({\mathbf x},t) &=& \Big[ - D({\mathbf x}_\parallel,t)
\, \delta'(x_3-q(t)) \, {\dot q}(t) \nonumber\\
&+&  \frac{\partial D({\mathbf x}_\parallel,t)}{\partial t} \,
\delta(x_3-q(t)) \Big] \mathbf{\hat e_3} \;,
\end{eqnarray}
where the presence of the last term is required by current conservation. 

A lengthier but otherwise entirely analogous calculation allows one to
obtain the spectral density for this case:
$$
{\mathcal P}(\omega) = \frac{1}{L^2}\, \vert\omega\vert \,
\int_0^{\omega/c}dk_\parallel\, k_\parallel^3 \,
\sqrt{(\frac{\omega}{c})^2-k_\parallel^2} \,
\int \frac{d\nu}{2\pi} \frac{d\nu'}{2\pi} 
$$
\begin{equation}\label{eq:despt}
\times \,{\tilde q}(\nu) \, {\widetilde D}({\mathbf k_\parallel}, \omega - \nu) 
{\widetilde D}(-{\mathbf k_\parallel}, -\omega - \nu') 
\,  {\tilde q}(\nu') \;,
\end{equation}
in terms of the space and time Fourier transforms on the patch potentials \cite{aclar}.

The derivation was performed without any assumption about the origin of the time
dependence. Let us now focus on the case in which the time dependences are
correlated. In the absence of external agents producing that time
dependence, it is reasonable to assume that they only depend on the time
difference between the two potentials. In Fourier space, that amounts to: 
\begin{equation}
\frac{1}{L^2} \, {\widetilde D}({\mathbf k_\parallel}, \omega)  
{\widetilde D}(-{\mathbf k_\parallel}, \omega') 
\to \widetilde{\Omega}({\mathbf k_\parallel},\omega) \, (2\pi)
\delta(\omega + \omega') \;,
\end{equation}
which inserted into (\ref{eq:despt}) yields:
\begin{eqnarray}
{\mathcal P}(\omega) &=& \frac{1}{L^2}\, \vert\omega\vert \,
\int_0^{\omega/c}dk_\parallel\, k_\parallel^3 \,
\sqrt{(\frac{\omega}{c})^2-k_\parallel^2} \nonumber\\
&\times& 
\int \frac{d\nu}{2\pi} 
{\tilde q}(-\nu) \, {\widetilde \Omega}({\mathbf k_\parallel}, \omega - \nu) 
\,  {\tilde q}(\nu) \;.
\end{eqnarray}

The last equation reduces to the static one for instantaneous correlation,
namely, when 
\begin{equation}
{\widetilde \Omega}({\mathbf k_\parallel}, \omega) \,=\, {\widetilde \Omega}({\mathbf k_\parallel})
(2\pi) \delta(\omega) \;.
\end{equation}

The calculation for time dependent dipole-layers could be a useful starting point
to develop a microscopic approach of the DCE. One should consider both 
electric and magnetic dipoles
as sources, with a particular time-dependent correlation function to describe
the quantum fluctuations.

\section{Examples and comparison with the DCE}\label{sec:examples}

As a first example of a patch potential autocorrelation, we first consider
the Gaussian approximation to the quasilocal correlation function proposed
in Ref. \cite{Behunin1}:
\begin{equation}
{\widetilde \Omega}({\mathbf k}_\parallel)=
\frac{\pi}{8}V_{rms}^2\ell^2 \exp[-\frac{1}{16}\vert  {\mathbf
k}_\parallel\vert^2 \ell^2 ] \;,
\end{equation}
where $V_{rms}$ is the variance of the potential and $\ell$ a characteristic length.
For this particular correlation, the spectral density reads
\begin{equation}
{\mathcal P}({\omega})= \, V_{rms}^2 \vert \tilde q(\omega)\vert^2  \frac{\omega^4}{
c^3} f\left(\ell\omega/4c\right)
\;,
\end{equation}
with
\begin{equation}
f(x)= \, \frac{2\pi}{x^3} \left[3x-(3+2x^2)D_+(x)\right]
\;,
\end{equation}
where  $D_+$ denotes the Dawson function (note that in the equations above
we have included the factor 4 coming from the image dipole layer).  For a
fixed frequency, the spectral density is a non-monotonous function of the
correlation length $\ell$. Indeed, ${\mathcal P}$ vanishes for $\ell\to 0$
(no patch potentials) and also vanishes in the opposite limit
$\ell\to\infty$, since by a simple application of symmetry arguments and Gauss' law one sees that
a uniform density of charges or dipoles
on a plane cannot radiate. Therefore, it must have a maximum at an
intermediate value. A plot of the function $f(x)$ shows that the maximum is
located at $x\sim 1.5$. As a consequence,  if the plane moves with a
definite frequency $\omega_0$, the radiation emitted is maximum when
the characteristic size of the patches is $\ell\sim 6 c/\omega_0$. 

As a second example, we will consider the sharp-cutoff model proposed in Ref.\cite{Speake}:
\begin{equation}
{\widetilde\Omega}({\mathbf k}_\parallel)= \frac{4\pi
V^2_{rms}}{k_{max}^2-k_{min}^2}\theta(\vert{\mathbf
k}_\parallel\vert-k_{min})\theta(k_{max}-\vert{\mathbf k}_\parallel\vert)
\;,
\label{VZ}
\end{equation}
which yields for the spectral density:
\begin{eqnarray}
{\mathcal P}({\omega})&=& \frac{64\pi V^2_{rms}}{15c^5(k_{max}^2-k_{min}^2)}\omega^6\vert\tilde q(\omega)\vert^2\nonumber\\
&\times&
\big[1-(\frac{k_{min}c}{\omega})^2\big]^{3/2}\big[2+3(\frac{k_{min}c}{\omega})^2\big]\, ,
\end{eqnarray}
 where we assumed that $\frac{k_{min}c}{\omega}<1$ and
$\frac{k_{max}c}{\omega}>1$. Note that ${\mathcal P}({\omega})$  vanishes 
 for  $\frac{k_{min}c}{\omega}>1$. As a consequence, 
 for the particular case in which the plane moves with a definite 
frequency $\omega_0$, there is a threshold to have a non vanishing emitted radiation  $k_{min}<\omega_0/c$.

In both examples one can consider the 
limiting case of small correlation length $\ell\omega/c\ll 1$, which is physically 
the more relevant limit.
For the quasilocal correlations we  obtain
\begin{equation}
{\mathcal P}({\omega})\simeq \, \frac{\pi}{60}V_{rms}^2\ell^2\frac{\omega^6}{c^5}
\vert \tilde q(\omega)\vert^2\, .
\end{equation}
A similar expression can be obtained for the sharp-cutoff model.

We will now compare the last result with that coming form the DCE.  A
single accelerated perfect mirror produces photons due to the interaction
with the quantum fluctuations of the electromagnetic field. On dimensional
grounds, in the non relativistic limit we expect the dissipative force per unit 
length on the mirror  ($f_{DCE}$) to be proportional to $\hbar\dddot a/c^4$  . This corresponds 
to a spectral 
density  proportional to
$\hbar\omega^6\vert \tilde q(\omega)\vert^2/c^4$. An explicit calculation yields~\cite{reviews dce}
\begin{equation}
{\mathcal P}_{DCE}({\omega})= \, \frac{\hbar}{30\pi^2}\frac{\omega^6}{c^4}
\vert \tilde q(\omega)\vert^2\, .
\end{equation}

Defining $\xi={\mathcal P}/{\mathcal P}_{DCE}=f_{rad}/f_{DCE}$, we obtain
\begin{equation}
\xi=\frac{\pi^3 V^2_{rms}\ell^2}{2\hbar c}\simeq  \frac{ V^2_{rms}}{(40\, mV)^2} \frac{\ell^2}{(100\, nm)^2}\, , 
\end{equation}
where we have written the result in terms of typical values that
characterize the patch potentials.  This shows that the reaction force due
to the classical radiation could be comparable or even larger than the one 
in the DCE. 

\bigskip

\section{Conclusions}\label{sec:conc}

We have shown that a real, accelerated metallic surface, produces classical
EM radiation due to the unavoidable presence of patch potentials. 
The calculation of the total emission spectrum for a flat surface
undergoing bounded motion is a rather straightforward exercise in classical
electrodynamics, and the result depends 
only on the most relevant physical quantity characterizing the patch potentials:
their autocorrelation function.

Remarkably, when the correlation length of the  patch potentials is
sufficiently small, the emitted radiation coincides with that of a single
moving dipole, and has the same frequency-dependence than the radiation
induced by the motion of a perfect conductor through the quantum vacuum.
Although these facts could have been anticipated by dimensional analysis,
the explicit calculations in this paper allowed us to compare the classical
radiation of the moving patches with the quantum radiation associated to
the DCE. Depending on the characteristics of the patches, the dissipative
effects associated to the classical radiation could be comparable with
those coming from the quantum vacuum for perfect conductors.  

We have considered the simplest situation, corresponding to a single
accelerated mirror. It is well known that the dynamical Casimir effect for
this configuration is far from being optimal regarding the possibility of
its experimental detection. It would be interesting to assess whether
classical radiation from patch potential mask the dynamical Casimir
effect or not in more realistic experimental settings, like for example a
closed cavity with variable length, in a regime of parametric
resonance~\cite{reviews dce}.

\section*{Acknowledgements}
We would  like to thank D. Dalvit for useful comments. This work was supported by ANPCyT, CONICET, and UNCuyo.


\begin{thebibliography}{99}
\bibitem{aclar} Since we are mainly interested in the radiation produced by the motion of the dipole layer, in this formula we omitted 
the contribution coming from a static layer,  that is present when the dipole density $D$ is time dependent.
\end{thebibliography}
\end{document}